\documentclass[graybox]{svmult}

\usepackage{mathptmx}       
\usepackage{helvet}         
\usepackage{courier}        
\usepackage{type1cm}        
\usepackage{makeidx}         
\usepackage{graphicx}        
\usepackage{multicol}        

\makeindex             

\begin{document}

\title*{Chiral Magnetic Effect on the Lattice}

\author{Arata Yamamoto}

\institute{Arata Yamamoto 
\at Quantum Hadron Physics Laboratory, RIKEN Nishina Center,\\
2-1 Hirosawa, Wako, Saitama 351-0198, Japan\\
\email{arayamamoto@riken.jp} }

\maketitle

\abstract{
We review recent progress on the lattice simulations of the chiral magnetic effect.
There are two different approaches to analyze the chiral magnetic effect on the lattice.
In one approach, the charge density distribution or the current fluctuation is measured under a topological background of the gluon field.
In the other approach, the topological effect is mimicked by the chiral chemical potential, and the induced current is directly measured.
Both approaches are now developing toward the exact analysis of the chiral magnetic effect.
}

\section{Introduction}

In the strong interaction, the gauge field forms nontrivial topology.
The existence of the topology has been theoretically established, while its observation is difficult in experiments.
{\it The chiral magnetic effect} is a possible candidate to detect the topological structure in heavy-ion collisions \cite{Kharzeev:2007jp}.
The chiral magnetic effect is the generation of an electric current in a strong magnetic field.

The essence of the chiral magnetic effect is the imbalance of the chirality, i.e., the number difference between the right-handed and left-handed quarks.
The magnetic field induces the electric currents of the right-handed and left-handed quarks in opposite directions.
If the chirality is imbalanced, a nonzero net electric current is induced. 
In a local domain of the QCD vacuum, the chiral imbalance is generated by the topological fluctuation and the axial anomaly.
In the global QCD vacuum, the chirality is balanced as a whole.
The strong theta parameter is experimentally zero, $\theta = 0$, although its reason is unknown.
This is {\it the strong CP problem}.
The chiral magnetic effect is regarded as the local violation of the CP symmetry.

Experimental facilities tried to measure the chiral magnetic effect through charged-particle correlations \cite{Abelev:2009ac,Abelev:2009ad}.
However, the interpretation of the experimental data is not yet conclusive.
On the theoretical side, the chiral magnetic effect has been studied in various frameworks, e.g., phenomenological models, the gauge-gravity duality, etc.
The chiral magnetic effect has been also studied in the lattice simulations.
The lattice simulation is a powerful framework to solve QCD nonperturbatively on computers.
By means of the lattice simulation, we can study the chiral magnetic effect from first principles in QCD.

\begin{figure}[t]
\begin{center}
\includegraphics[scale=0.5]{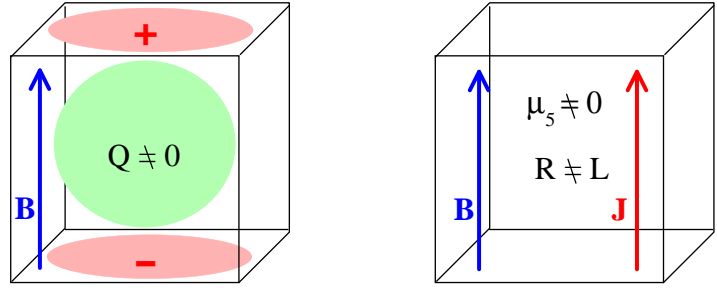}
\caption{A cartoon of how to observe the chiral magnetic effect on the lattice.
Left: a topological charge $Q$ of the gauge field induces a nonuniform current density distribution.
Right: a chiral chemical potential $\mu_5$ induces a uniform electric current.}
\label{fig1}
\end{center}
\end{figure}

There are two approaches to analyze the chiral magnetic effect in the lattice simulation.
In other words, there are two different ways to generate the chiral imbalance:
\begin{enumerate}
 \item topological charge \cite{Buividovich:2009zzb,Buividovich:2009wi,Buividovich:2009zj,Braguta:2010ej,Abramczyk:2009gb,Ishikawa:2011}
 \item chiral chemical potential \cite{Yamamoto:2011gk,Yamamoto:2011qa,Yamamoto:2011ks}
\end{enumerate}
These concepts are schematically depicted in Fig.~\ref{fig1}.
In the first case, a topological charge of the background gauge field generates the chiral imbalance, which is spatially nonuniform.
When an external magnetic field is applied, a current density distribution appears around the topological object.
In the second case, a chiral chemical potential generates the chirally imbalanced matter, which is spatially uniform.
A uniform electric current is induced by the external magnetic field.

In this chapter, we overview the theoretical background and the current status of the lattice studies of the chiral magnetic effect.
Here we focus only on the lattice aspect of the chiral magnetic effect.
For the theoretical and phenomenological aspects, see the corresponding chapters.
We use the Euclidean metric and the lattice unit in the following sections.

\section{Basics of the lattice simulation}

The basic formalism of the lattice simulation has been well established.
For the details, see the textbooks \cite{Creutz:1984mg,Montvay:1994cy,Smit:2002ug,Rothe:2005nw}.
The formalism is based on the Euclidean QCD partition function
\begin{eqnarray}
Z &=& \int DU D\bar{\psi} D\psi e^{-S_G[U]-S_F[\bar{\psi},\psi,U]} \nonumber \\
&=& \int DU \det D[U] e^{-S_G[U]} \ .
\end{eqnarray}
The space-time is discretized as a hypercubic lattice.
The gluon field is written as the SU(3) link variable
\begin{eqnarray}
U_\mu(x) = \exp (i g t^a A^a_\mu (x)) \ .
\end{eqnarray}
The functional integral is numerically evaluated by the Monte Carlo simulation.
We generate gauge configurations, which are sets of the link variable, and then calculate the expectation value of an operator as
\begin{eqnarray}
\langle O[U] \rangle = \frac{1}{N_{\rm conf}} \sum_{\rm \{ U \} } O[U] \ .
\end{eqnarray}
The gauge configurations are generated to satisfy the probability weight $P = \det D[U] e^{-S_G[U]} $.
The simulation including the fermion determinant is called {\it the dynamical QCD simulation} or {\it the full QCD simulation}.
{\it The quenched approximation} is often used to reduce the computational cost.
In the quenched approximation, the fermion determinant is ignored and the probability weight is  $P = e^{-S_G[U]} $.
The quenched gauge configurations are independent of the fermion action.

The probability weight must be positive real, otherwise it cannot be interpreted as the probability weight.
In QCD, the fermion action becomes complex at a finite quark chemical potential.
The definition of the probability weight must be modified, e.g., by the reweighting method \cite{Ferrenberg:1988yz}.
Even after the modification, the Monte Carlo simulation severely suffers from strong sign fluctuation.
This is known as {\it the sign problem}.
The sign problem at the quark chemical potential is an important unsolved problem in the lattice simulation \cite{deForcrand:2010ys}.
As shown later, a chiral chemical potential does not cause the sign problem.
This is similar to two-color QCD \cite{Hands:1999md,Kogut:2001na,Kogut:2002tm} and an isospin chemical potential \cite{Kogut:2002tm,Kogut:2002zg,Kogut:2004zg,Nakamura:2003gj,deForcrand:2007uz}.

The basic observable of the chiral magnetic effect is the local vector current density
\begin{eqnarray}
j_\mu (x) = \bar{\psi}(x) \gamma_\mu \psi(x) \ .
\label{eqlvc}
\end{eqnarray}
The fourth (zeroth) component corresponds to the local charge density.
For calculating the local vector current density, we consider the Dirac eigenvalue problem
\begin{eqnarray}
D[U] \phi_k(x)  = (i\lambda_k + m) \phi_k(x) \ ,
\label{eqDEV}
\end{eqnarray}
and use the identity
\begin{eqnarray}
\langle \bar{\psi}(x) \gamma_\mu \psi(x) \rangle = 
\langle {\rm tr} \gamma_\mu D[U]^{-1} \rangle =
\left \langle \sum_k \frac{\bar{\phi}_k(x) \gamma_\mu \phi_k(x)}{i\lambda_k +m} \right \rangle \ .
\end{eqnarray}
Thus, its expectation value is obtained by inverting or diagonalizing the Dirac operator $D[U]$.
In the case of diagonalizing, we can calculate the local vector current density of each Dirac eigenmode.

Chiral symmetry is a nontrivial problem on the lattice due to the Nielsen-Ninomiya no-go theorem \cite{Nielsen:1980rz,Nielsen:1981xu}.
Most lattice fermions more or less break chiral symmetry.
The lattice fermion with exact chiral symmetry has been known, although its computational cost is rather large in the dynamical simulation.
We should select an appropriate lattice fermion, corresponding to the purpose of the simulation.
For the details of the lattice fermions and chiral symmetry, see the reviews \cite{Creutz:2000bs,Neuberger:2001nb,Chandrasekharan:2004cn}.

For a magnetic field, the QED gauge field is also introduced.
When the magnetic field is external, i.e., not dynamical, the QED field strength term does not exist in the action.
To couple the fermions to the magnetic field, the Dirac operator is replaced as
\begin{eqnarray}
D[U] \to D[u U]
\end{eqnarray}
with the U(1) link variable
\begin{eqnarray}
u_\mu(x) = \exp(i q A_\mu(x)) \ .
\end{eqnarray}
We can apply a homogeneous magnetic field in a finite-volume box with periodic boundary conditions.
For the homogeneous magnetic field in the $z$-direction, the U(1) link variables are set as
\begin{eqnarray}
u_1(x) &=& \exp(-i q B N_s y) \quad {\rm at}\ x=N_s  \\
u_2(x) &=& \exp(i q B x) \\
u_\mu(x) &=& 1 \quad {\rm for}\ {\rm other}\ {\rm components}
\end{eqnarray}
in the lattice volume  $N_s^3 \times N_t$ \cite{AlHashimi:2008hr}.
In this setup, the magnetic field is quantized as
\begin{eqnarray}
q B = \frac{2\pi}{N_s^2} \times {\rm (integer)} \ .
\end{eqnarray}
This integer is the input parameter which controls the strength of the magnetic field in the simulation.

\section{Lattice simulation with a topological background}

The gauge configuration possesses a topological charge.
The topological charge of the gauge configuration is given as
\begin{eqnarray}
Q = \frac{g^2}{64\pi^2} \int d^4 x \ \epsilon_{\mu \nu \lambda \rho} F^a_{\mu \nu}(x) F^a_{\lambda \rho}(x) \ .
\end{eqnarray}
Euclidean topological objects, such as the instanton, can be reproduced on the lattice when the gauge configuration is smooth enough \cite{Teper:1999wp}.

The fermion feels the background topology of the gauge configuration through {\it the zero mode}.
The zero mode is defined as the eigenmode $\phi_k$ which has the zero eigenvalue $i\lambda_k = 0$ in Eq.~(\ref{eqDEV}).
The topological charge and the zero mode are related through the Atiyah-Singer index theorem
\begin{eqnarray}
N_R - N_L = N_f Q \ ,
\end{eqnarray}
where $N_R$ and $N_L$ are the numbers of the right-handed zero modes and the left-handed zero modes, respectively \cite{Atiyah:1968}.
The fermion zero mode is essential to generate the chiral imbalance in the topological background.
We must use the lattice fermion which is sensitive to the zero mode and satisfies the index theorem, e.g., the overlap fermion.

Naively, it is impossible to measure the local vector current density (\ref{eqlvc}) in the topological background.
The reason is as follows.
In the QCD vacuum, the positive and negative topological charges appear with the same probability.
In the  simulation, the numbers of the gauge configurations with the positive and negative topological charges are the same, namely
\begin{eqnarray}
\langle Q \rangle = 0 \ .
\end{eqnarray}
The positive and negative topological charges induce the vector current in opposite directions.
When all the Monte Carlo samples are averaged in all the topological sectors, the net vector current is zero.
To measure the vector current, we must fix the topological sector by the lattice action which suppresses topology changing transitions \cite{Fukaya:2006vs}.
Although the fixed-topology simulation cannot reproduce the $\theta = 0$ vacuum, we can obtain a finite expectation value of the vector current.

\begin{figure}[h]
\begin{minipage}{0.5\hsize}
\begin{center}
\includegraphics[scale=0.15]{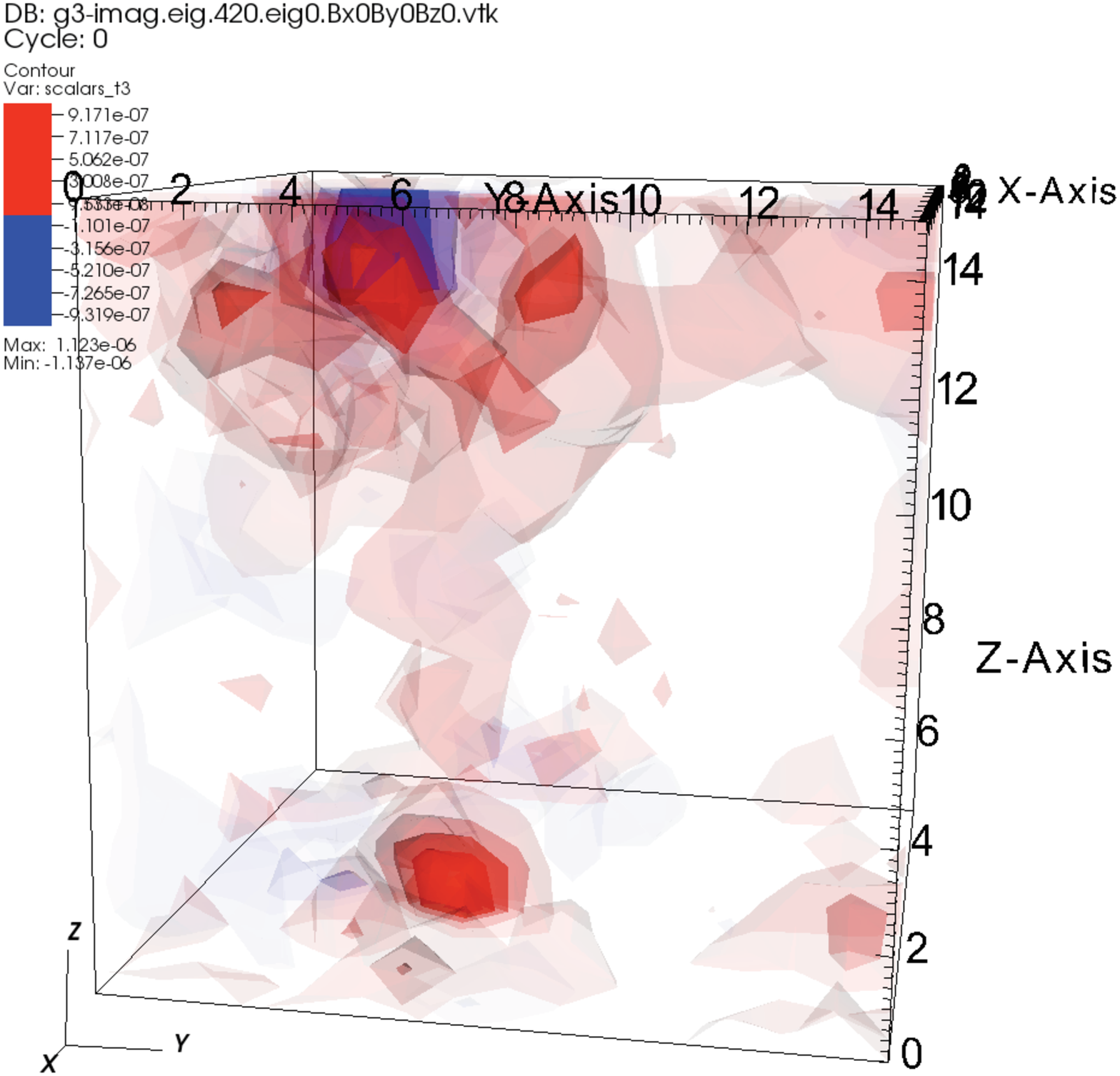}
\end{center}
\end{minipage}
\begin{minipage}{0.5\hsize}
\begin{center}
\includegraphics[scale=0.15]{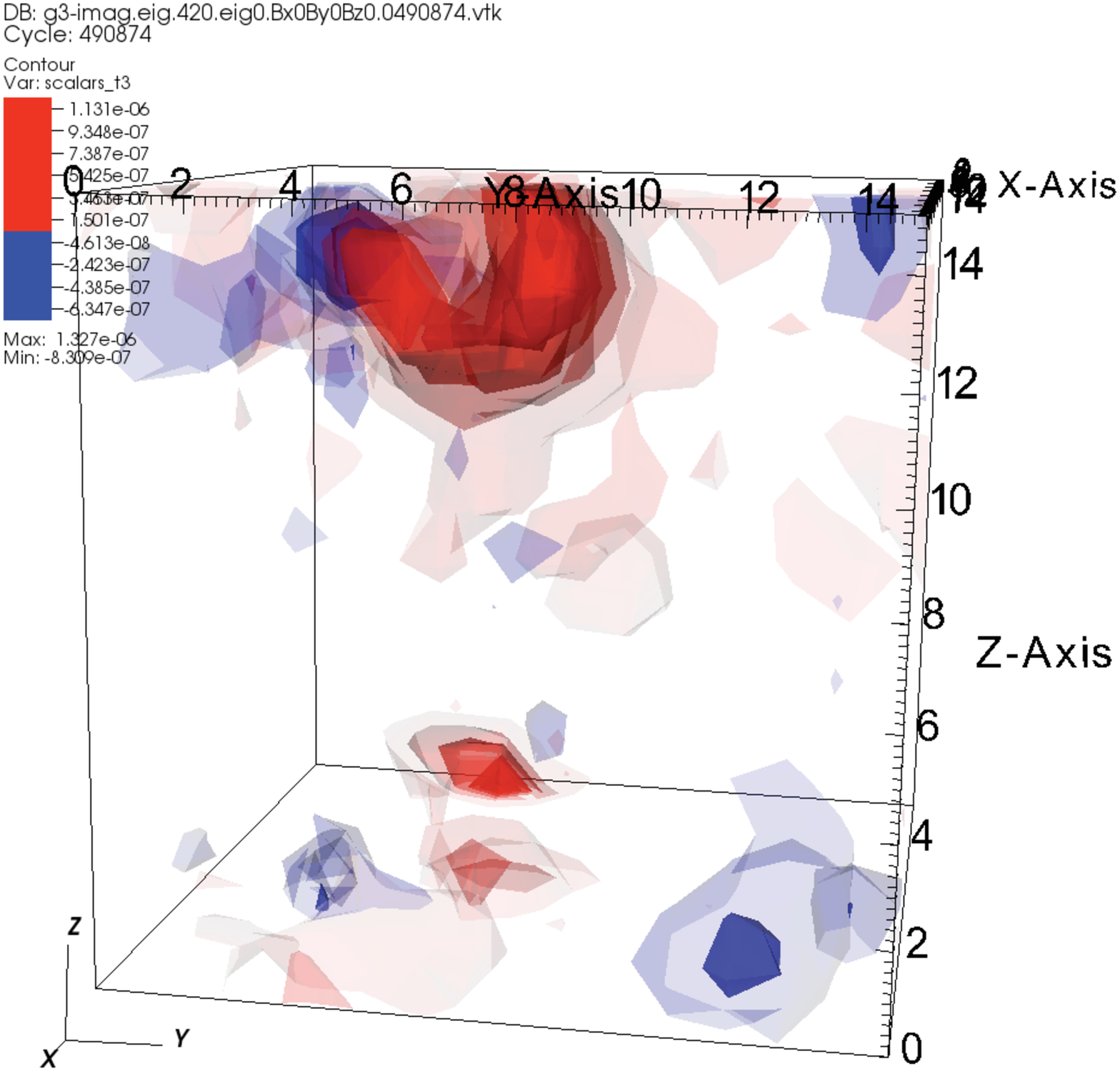}
\end{center}
\end{minipage}
\caption{The charge density distribution $\rho(x)$ in the (2+1)-flavor dynamical QCD+QED simulation at $a^2qB = 0$ (left) and 0.0490874 (right) \cite{Abramczyk:2009gb}.
The magnetic field is applied along the $z$-axis.
The temperature is above the critical temperature.
}
\label{fig2}
\end{figure}

The fixed-topology analysis has been done in the (2+1)-flavor dynamical QCD simulation with the domain-wall fermion \cite{Abramczyk:2009gb,Ishikawa:2011}.
This simulation includes not only the external magnetic field but also the dynamical QED effect.
The domain-wall fermion does not have the exact zero mode due to small explicit chiral symmetry breaking, but has the ``near'' zero mode which becomes the exact zero mode in an ideal limit.
In Fig.~\ref{fig2}, we show the charge density distribution of one near zero mode in one typical gauge configuration \cite{Abramczyk:2009gb}.
The charge density of the $k$-th eigenmode is defined as
\begin{eqnarray}
\rho_k (x) = \frac{\phi_k^\dagger(x) \phi_k(x)}{i\lambda_k +m} \ .
\end{eqnarray}
The simulation was performed above the critical temperature.
The charge density distribution at $B\ne 0$ differs from that at $B=0$.
This result suggests that some relation exists between the charge density and the magnetic field.
However, the exact relation is not clear in this simulation.
We need further investigation for evidence of the chiral magnetic effect.

Except for the fixed-topology simulation, the vector current itself is zero because of the CP oddness.
In this case, a numerical observable is the CP-even quantity which reflects the topological fluctuation
\begin{eqnarray}
\frac{\langle Q^2 \rangle}{V} \simeq (200 \ {\rm MeV})^4 \ .
\end{eqnarray}
For instance, the fluctuation of the vector current is CP-even.
This situation is similar to the experimental observation.
An experimental observable must be CP-even, although the chiral magnetic effect is a CP-odd process.
We have to extract the CP-odd information from the CP-even particle correlation.
This kind of analysis is not easy because the fluctuation can be easily induced by other irrelevant effects.
The irrelevant contributions must be subtracted correctly.

\begin{figure}[h]
\begin{minipage}{0.5\hsize}
\begin{center}
\rotatebox{-90}{\includegraphics[scale=0.25]{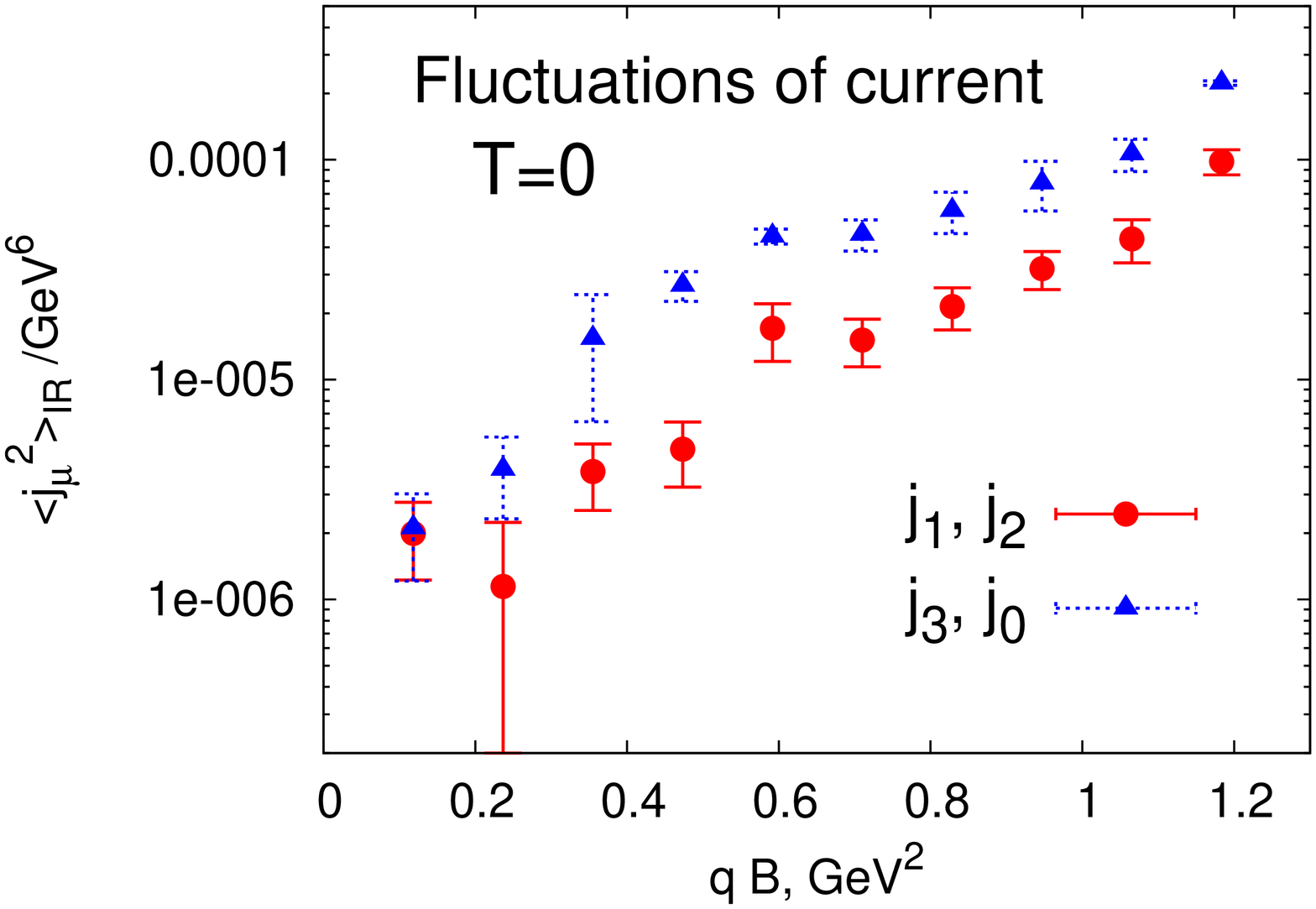}}
\end{center}
\end{minipage}
\begin{minipage}{0.5\hsize}
\begin{center}
\rotatebox{-90}{\includegraphics[scale=0.25]{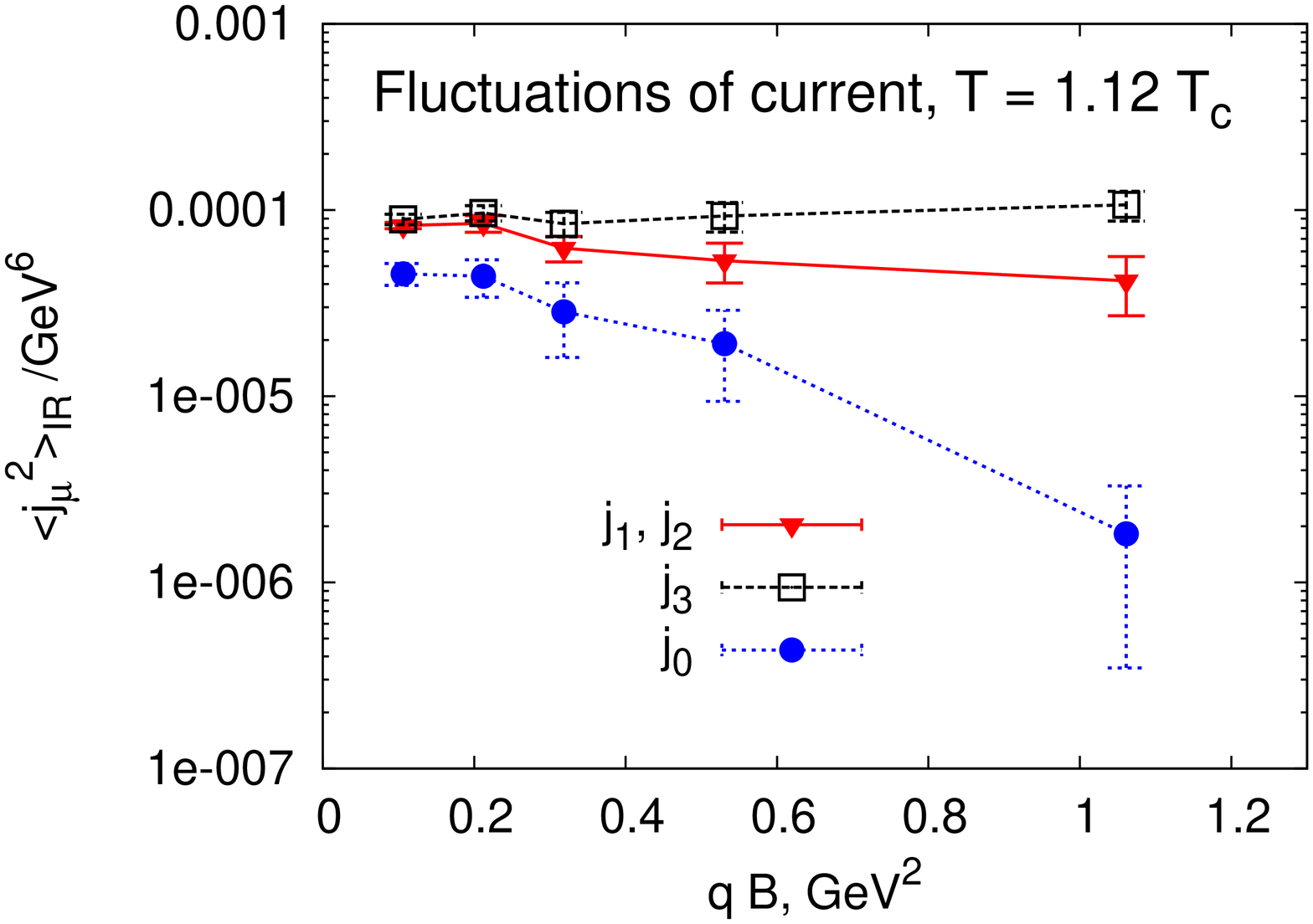}}
\end{center}
\end{minipage}
\caption{The current fluctuation $\langle j_\mu^2 \rangle_{\rm IR}$ in the quenched SU(2) simulation at $T=0$ (left) and $T=1.12T_c$ (right) \cite{Buividovich:2009wi}.
The magnetic field is applied in the $\mu=3$ direction.
}
\label{fig3}
\end{figure}

The fluctuation $\langle j_\mu^2 \rangle$ of the vector current was calculated in the quenched SU(2) simulation at zero temperature \cite{Buividovich:2009zzb}, in the quenched SU(2) simulation at finite temperature \cite{Buividovich:2009wi,Buividovich:2009zj}, and in the quenched SU(3) simulation \cite{Braguta:2010ej}.
The overlap Dirac operator was adopted in these simulations, although the zero modes were ignored.
The vector currents are zero in all the directions because the topological sector is not fixed, but the current fluctuation is nonzero.
In Fig.~\ref{fig3}, we show the current fluctuation in the quenched SU(2) simulation below and above the critical temperature \cite{Buividovich:2009wi}.
The ultraviolet part of the fluctuation is subtracted to obtain a clear signal as
\begin{eqnarray}
\langle j_\mu^2 \rangle_{\rm IR}
= \frac{1}{V} \sum_{\rm site} \langle j_\mu^2 (x) \rangle_{B,T}
- \frac{1}{V} \sum_{\rm site} \langle j_\mu^2 (x) \rangle_{B=0,T=0} \ ,
\end{eqnarray}
where the index $\mu$ is not summed over.
At zero temperature $T = 0$, all the fluctuations grow at stronger magnetic field.
In particular, the longitudinal fluctuation $\langle j_3^2 \rangle$ grows faster than transverse fluctuations $\langle j_1^2 \rangle = \langle j_2^2 \rangle$.
Above the critical temperature $T> T_c$, the longitudinal fluctuation is insensitive and the transverse fluctuations decrease at stronger magnetic field.
As a consequence, the ratio of the longitudinal fluctuation to the transverse fluctuation is enhanced by the magnetic field in both cases.

As shown above, the magnetic field affects the charge density distribution and the current fluctuation.
Note however that we must carefully check whether its origin is actually the chiral magnetic effect.
In general, a strong magnetic field can induce a strong current fluctuation in the longitudinal direction, even if there is no topological object.
This complication is the same as that in experiments.
For identifying the chiral magnetic effect, we must distinguish a small topological contribution from other large contaminations in a high-precision simulation.

\section{Lattice simulation with a chiral chemical potential}

Another possible source of the chiral imbalance is a chiral chemical potential.
The chiral chemical potential $\mu_5$ is defined as
\begin{eqnarray}
D(\mu_5) = \gamma_\mu (\partial_\mu +i g t^a A^a_\mu(x)) + m + \mu_5 \gamma_4 \gamma_5
\label{eqDana}
\end{eqnarray}
in the continuum space \cite{Fukushima:2008xe}.
The chiral chemical potential directly couples to the chiral charge
\begin{eqnarray}
N_5 \equiv N_R - N_L = - \int d^3x \langle \bar{\psi}(x) \gamma_4 \gamma_5 \psi(x) \rangle  \ .
\end{eqnarray}
By using the chiral chemical potential, we can generate a chirally imbalanced QCD matter in equilibrium.
The chiral chemical potential is the external parameter which tunes the chiral charge instead of the topological charge.
The chiral chemical potential does not exist in the original QCD action because the chiral charge is not a conserved quantity.
It is not a ``chemical potential'' in the exact sense.

Because the topological charge is not necessary in this approach, the sensitivity to the zero mode is not important for the choice of the fermion action.
For example, the lattice Dirac operator of the Wilson fermion is
\begin{eqnarray}
\frac{1}{m} D_{\rm W}(\mu_5) &=& 1
- \kappa \sum_i \bigl[ (1-\gamma_i)T_{i+} \nonumber + (1+\gamma_i)T_{i-} \bigr] \nonumber\\
&&- \kappa \bigl[ (1-\gamma_4 e^{\mu_5\gamma_5})T_{4+} + (1+\gamma_4 e^{-\mu_5\gamma_5})T_{4-} \bigr] \ ,
\label{eqDlat}
\end{eqnarray}
with
\begin{eqnarray}
&& \kappa \equiv \frac{1}{2m + 8} \\
&&[T_{\mu +}]_{x,y} \equiv U_\mu (x) \delta_{x+\hat{\mu},y} \\
&&[T_{\mu -}]_{x,y} \equiv U^\dagger_\mu (y) \delta_{x-\hat{\mu},y} \ .
\end{eqnarray}
The chiral chemical potential is introduced as the exponential matrix factor
\begin{eqnarray}
e^{\pm \mu_5\gamma_5} = \cosh \mu_5 \pm \gamma_5\sinh \mu_5 \ ,
\end{eqnarray}
which is the straightforward analogy to a quark chemical potential \cite{Hasenfratz:1983ba}.
The Wilson-Dirac operator (\ref{eqDlat}) reproduces the continuum form (\ref{eqDana}) in the continuum limit.

A notable feature of the chiral chemical potential is that it does not cause the sign problem unlike the quark chemical potential.
The Wilson-Dirac operator (\ref{eqDlat}) is ``$\gamma_5$-Hermitian'',
\begin{eqnarray}
\gamma_5 D(\mu_5) = [ \gamma_5 D (\mu_5) ]^\dagger
\quad {\rm or} \quad
\gamma_5 D(\mu_5) \gamma_5 =  D^\dagger (\mu_5)\ .
\end{eqnarray}
In the two-flavor case, the fermion determinant is positive real,
\begin{equation}
\det \left(
\begin{array}{cc}
D(\mu_5) & 0 \\
0 & D(\mu_5)
\end{array}
  \right) 
= \det D(\mu_5)  \det \gamma_5 D(\mu_5) \gamma_5
= |\det D(\mu_5)|^2 \ge 0 \ .
\end{equation}
Therefore there is no sign problem.
We can exactly simulate a kind of finite density QCD matter by the chiral chemical potential.

\begin{figure}[h]
\begin{center}
\includegraphics[scale=1]{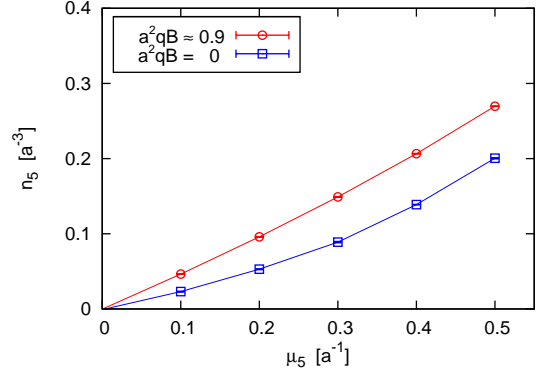}
\caption{The chiral charge density $n_5$ in the two-flavor dynamical QCD simulation \cite{Yamamoto:2011ks}.
The lattice spacing is $a \simeq 0.13$ fm and the temperature is $T \simeq 400$ MeV.}
\label{fig4}
\end{center}
\end{figure}

In Fig.~\ref{fig4}, we show the chiral charge density
\begin{eqnarray}
n_5 = \frac{N_5}{V} = - \frac{1}{V} \sum_{\rm site} \langle \bar{\psi}(x) \gamma_4\gamma_5 \psi(x) \rangle
\end{eqnarray}
of the Wilson fermion in the two-flavor dynamical QCD simulation \cite{Yamamoto:2011ks}.
The lattice spacing is $a \simeq 0.13$ fm.
The physical temperature is $T \simeq 400$ MeV, which is above the critical temperature.
The chiral charge density is finite at a finite chiral chemical potential.
This means that the uniform chirally imbalanced matter is realized on the lattice.
The total chiral charge in this lattice volume is $N_5 = n_5V \simeq O(10^3)$.
This number is much larger than a typical number of the topological charge.
The typical number of the topological charge is $O(10)$ at most in the conventional lattice simulation.
Owing to the large chiral imbalance, the analysis of the chiral magnetic effect becomes easy.

\begin{figure}[h]
\begin{minipage}{0.5\hsize}
\begin{center}
\includegraphics[scale=0.8]{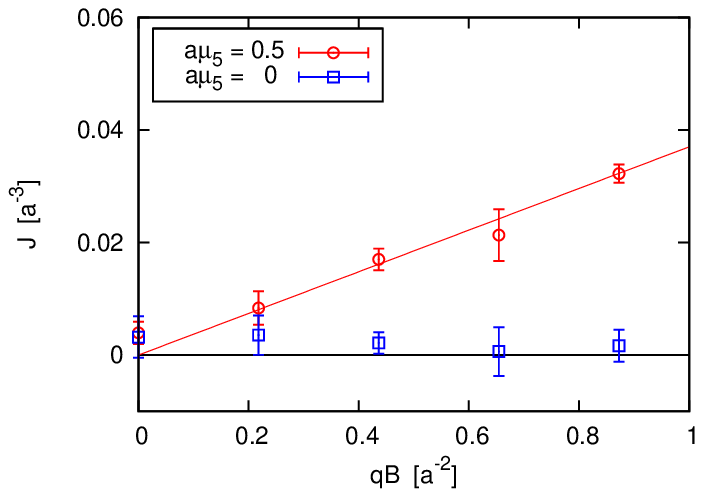}
\end{center}
\end{minipage}
\begin{minipage}{0.5\hsize}
\begin{center}
\includegraphics[scale=0.8]{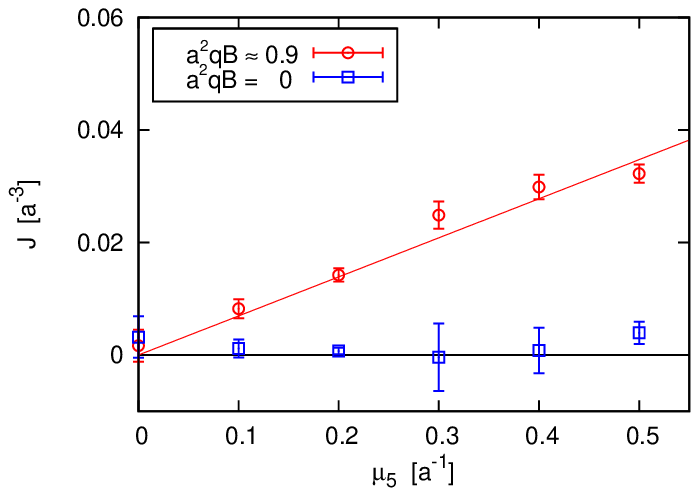}
\end{center}
\end{minipage}
\caption{The vector current density $J$ in the two-flavor dynamical QCD simulation \cite{Yamamoto:2011ks}.
The data are plotted as a function of the magnetic field $B$ (left) and of the chiral chemical potential $\mu_5$ (right).
The lattice spacing is $a \simeq 0.13$ fm and the temperature is $T \simeq 400$ MeV.
}
\label{fig5}
\end{figure}

For the analysis of the chiral magnetic effect, the local vector current density (\ref{eqlvc}) was measured.
The vector current is induced only in the longitudinal direction of the magnetic field.
The transverse components are exactly zero, $\langle j_1 \rangle = \langle j_2 \rangle = 0$.
In Fig.~\ref{fig5}, the induced current 
\begin{eqnarray}
J =  \frac{1}{V} \sum_{\rm site} \langle j_3(x) \rangle
\end{eqnarray}
is plotted as a function of the magnetic field and of the chiral chemical potential.
This induced current is direct evidence of the chiral magnetic effect.
The induced current is a linearly increasing function in both cases.
Therefore, the functional form is
\begin{eqnarray}
J = N_{\rm dof} C \mu_5 qB \ .
\label{eqJlat}
\end{eqnarray}
Because all the fermions have the same charge in this simulation, the prefactor is $N_{\rm dof} = 3 ({\rm color}) \times 2 ({\rm flavor}) =6$.
The overall coefficient $C$ characterizes the strength of the induced current.
This functional form is consistent with the analytical formula,
\begin{eqnarray}
J = N_{\rm dof} \frac{1}{2\pi^2} \mu_5 qB \ ,
\label{eqJana}
\end{eqnarray}
which was derived from the Dirac equation coupled with the background magnetic field \cite{Fukushima:2008xe}.
Note that Eq.~(\ref{eqJana}) is different from Ref.~\cite{Fukushima:2008xe} by $q$ due to the definition of the electric current, i.e., $J_{\rm EM} = q J$.
If there are several fermions with different charges, the total electric current is $J_{\rm EM} = \sum_i q_i J_i = \sum_i q_i^2 C \mu_5 B$.

The overall coefficient is $C = 0.013 \pm 0.001$ in this lattice simulation and $C = 1/(2\pi^2) \simeq 0.05$ in the analytical formula.
The induced current seems somehow smaller than the analytical formula.
However, these overall coefficients should not be compared naively.
For the quantitative argument, it is necessary to estimate several systematic effects in the lattice simulation.
One important effect is the renormalization of the local vector current.
The local vector current (\ref{eqlvc}) is not renormalization-group invariant on the lattice \cite{Karsten:1980wd}.
This property is different from that in the continuum theory.
The local vector current is renormalization-group invariant in the continuum theory because of the Ward identity.
We must take the continuum limit to compare the induced currents on the lattice and in the continuum.
By taking the continuum limit, we can also remove other lattice discretization artifacts.
For example, the Wilson fermion explicitly breaks the chiral symmetry due to the discretization artifact.

The systematic errors were partly estimated in Ref.~\cite{Yamamoto:2011ks}.
By varying simulation parameters, the dependences of the overall coefficient were examined in the quenched simulation.
Although the dynamical QCD simulation is necessary for the quantitative argument, the quenched simulation is useful to understand which systematic effect is important.
Actually, the quenched results were qualitatively similar to the dynamical QCD results.
It turned out that the induced current is insensitive to the temperature, the quark mass, and the spatial volume.
However, the overall coefficient strongly depends on the lattice spacing.
The overall coefficient increases near the continuum limit.
This systematic analysis indicates that the continuum extrapolation is necessary for the quantitative argument.

Another important effect is chiral symmetry.
It is difficult to discuss chiral symmetry using the naive Wilson fermion.
The Wilson fermion explicitly breaks chiral symmetry at a finite lattice spacing, while the explicit breaking vanishes in the continuum limit.
One possible origin of the strong lattice spacing dependence might be this artificial chiral symmetry breaking.
We should investigate the role of chiral symmetry in the chiral magnetic effect by performing the same analysis with a chiral lattice fermion, such as the domain-wall fermion or the overlap fermion.

\section{Conclusion}

In this review, we have overviewed the lattice studies of the chiral magnetic effect.
The vector current and its fluctuation were measured in the chiral imbalance, which is generated by the topological charge or the chiral chemical potential.
We should develop these pioneering works in future.
In the future works, it is important to respect the essential pieces of the chiral magnetic effect, in particular, the fermion zero mode and chiral symmetry.

We see that the chiral magnetic effect is an observable phenomenon on the lattice.
The lattice simulation is a hopeful approach to study the chiral magnetic effect in ``numerical'' experiments.

\section*{Acknowledgments}

The author is supported by the Special Postdoctoral Research Program of RIKEN.


\begin{thebibliography}{99}

\bibitem{Kharzeev:2007jp} 
  D.~E.~Kharzeev, L.~D.~McLerran and H.~J.~Warringa,
  Nucl.\ Phys.\ A {\bf 803}, 227 (2008)  [arXiv:0711.0950 [hep-ph]].

\bibitem{Abelev:2009ac} 
  B.~I.~Abelev {\it et al.}  [STAR Collaboration],
  Phys.\ Rev.\ Lett.\  {\bf 103}, 251601 (2009)  [arXiv:0909.1739 [nucl-ex]].

\bibitem{Abelev:2009ad} 
  B.~I.~Abelev {\it et al.}  [STAR Collaboration],
  Phys.\ Rev.\ C {\bf 81}, 054908 (2010)  [arXiv:0909.1717 [nucl-ex]].

\bibitem{Buividovich:2009zzb}
  P.~V.~Buividovich, E.~V.~Luschevskaya, M.~I.~Polikarpov and M.~N.~Chernodub,
  JETP Lett.\  {\bf 90}, 412 (2009);
  Pisma Zh.\ Eksp.\ Teor.\ Fiz.\  {\bf 90}, 456 (2009).

\bibitem{Buividovich:2009wi} 
  P.~V.~Buividovich, M.~N.~Chernodub, E.~V.~Luschevskaya and M.~I.~Polikarpov,
  Phys.\ Rev.\ D {\bf 80}, 054503 (2009)  [arXiv:0907.0494 [hep-lat]].

\bibitem{Buividovich:2009zj} 
  P.~V.~Buividovich, M.~N.~Chernodub, E.~V.~Luschevskaya and M.~I.~Polikarpov,
  PoS {\bf LAT2009}, 080 (2009)  [arXiv:0910.4682 [hep-lat]].

\bibitem{Braguta:2010ej} 
  V.~V.~Braguta, P.~V.~Buividovich, T.~Kalaydzhyan, S.~V.~Kuznetsov and M.~I.~Polikarpov,
  PoS {\bf LATTICE2010}, 190 (2010)  [arXiv:1011.3795 [hep-lat]].

\bibitem{Abramczyk:2009gb} 
  M.~Abramczyk, T.~Blum, G.~Petropoulos and R.~Zhou,
  PoS {\bf LAT2009}, 181 (2009)  [arXiv:0911.1348 [hep-lat]].

\bibitem{Ishikawa:2011}
  T.~Ishikawa and T.~Blum,
  PoS {\bf LATTICE2011}, 196 (2011).

\bibitem{Yamamoto:2011gk} 
  A.~Yamamoto,
  Phys.\ Rev.\ Lett.\  {\bf 107}, 031601 (2011)  [arXiv:1105.0385 [hep-lat]].

\bibitem{Yamamoto:2011qa} 
  A.~Yamamoto,
  PoS {\bf LATTICE2011}, 220 (2011)  [arXiv:1108.0937 [hep-lat]].

\bibitem{Yamamoto:2011ks} 
  A.~Yamamoto,
  Phys.\ Rev.\ D {\bf 84}, 114504 (2011)  [arXiv:1111.4681 [hep-lat]].

\bibitem{Creutz:1984mg} 
  M.~Creutz,
  {\it Quarks, Gluons And Lattices,}
  Cambridge, UK: Univ. Pr. (1983). (Cambridge Monographs on Mathematical Physics)

\bibitem{Montvay:1994cy} 
  I.~Montvay and G.~Munster,
  {\it Quantum fields on a lattice,}
  Cambridge, UK: Univ. Pr. (1994). (Cambridge Monographs on Mathematical Physics)

\bibitem{Smit:2002ug} 
  J.~Smit,
  {\it Introduction to quantum fields on a lattice: A robust mate,}
  Cambridge Lect.\ Notes Phys.\  {\bf 15}, 1 (2002).

\bibitem{Rothe:2005nw} 
  H.~J.~Rothe,
  {\it Lattice gauge theories: An Introduction,}
  World Sci.\ Lect.\ Notes Phys.\  {\bf 74}, 1 (2005).

\bibitem{Ferrenberg:1988yz} 
  A.~M.~Ferrenberg and R.~H.~Swendsen,
  Phys.\ Rev.\ Lett.\  {\bf 61}, 2635 (1988).

\bibitem{deForcrand:2010ys} 
  For a review, P.~de Forcrand,
  PoS {\bf LAT2009}, 010 (2009)  [arXiv:1005.0539 [hep-lat]].

\bibitem{Hands:1999md} 
  S.~Hands, J.~B.~Kogut, M.~-P.~Lombardo and S.~E.~Morrison,
  Nucl.\ Phys.\ B {\bf 558}, 327 (1999)  [hep-lat/9902034].

\bibitem{Kogut:2001na} 
  J.~B.~Kogut, D.~K.~Sinclair, S.~J.~Hands and S.~E.~Morrison,
  Phys.\ Rev.\ D {\bf 64}, 094505 (2001)  [hep-lat/0105026].

\bibitem{Kogut:2002tm} 
  J.~B.~Kogut and D.~K.~Sinclair,
  Phys.\ Rev.\ D {\bf 66}, 014508 (2002)  [hep-lat/0201017].

\bibitem{Kogut:2002zg} 
  J.~B.~Kogut and D.~K.~Sinclair,
  Phys.\ Rev.\ D {\bf 66}, 034505 (2002)  [hep-lat/0202028].

\bibitem{Kogut:2004zg} 
  J.~B.~Kogut and D.~K.~Sinclair,
  Phys.\ Rev.\ D {\bf 70}, 094501 (2004)  [hep-lat/0407027].

\bibitem{Nakamura:2003gj} 
  A.~Nakamura and T.~Takaishi,
  Nucl.\ Phys.\ Proc.\ Suppl.\  {\bf 129}, 629 (2004)  [hep-lat/0310052].

\bibitem{deForcrand:2007uz} 
  P.~de Forcrand, M.~A.~Stephanov and U.~Wenger,
  PoS {\bf LAT2007}, 237 (2007)  [arXiv:0711.0023 [hep-lat]].

\bibitem{Nielsen:1980rz} 
  H.~B.~Nielsen and M.~Ninomiya,
  Nucl.\ Phys.\ B {\bf 185}, 20 (1981);
  Erratum-ibid \ B {\bf 195}, 541 (1982).

\bibitem{Nielsen:1981xu} 
  H.~B.~Nielsen and M.~Ninomiya,
  Nucl.\ Phys.\ B {\bf 193}, 173 (1981).

\bibitem{Creutz:2000bs} 
  M.~Creutz,
  Rev.\ Mod.\ Phys.\  {\bf 73}, 119 (2001)  [hep-lat/0007032].

\bibitem{Neuberger:2001nb} 
  H.~Neuberger,
  Ann.\ Rev.\ Nucl.\ Part.\ Sci.\  {\bf 51}, 23 (2001)  [hep-lat/0101006].

\bibitem{Chandrasekharan:2004cn} 
  S.~Chandrasekharan and U.~J.~Wiese,
  Prog.\ Part.\ Nucl.\ Phys.\  {\bf 53}, 373 (2004)  [hep-lat/0405024].

\bibitem{AlHashimi:2008hr} 
  M.~H.~Al-Hashimi and U.~-J.~Wiese,
  Annals Phys.\  {\bf 324}, 343 (2009)  [arXiv:0807.0630 [quant-ph]].

\bibitem{Teper:1999wp} 
  For a review, M.~Teper,
  Nucl.\ Phys.\ Proc.\ Suppl.\  {\bf 83}, 146 (2000)  [hep-lat/9909124].

\bibitem{Atiyah:1968}
  M.~F.~Atiyah and I.~M.~Singer,
  Ann. Math. {\bf 87}, 484 (1968).

\bibitem{Fukaya:2006vs} 
  H.~Fukaya {\it et al.}  [JLQCD Collaboration],
  Phys.\ Rev.\ D {\bf 74}, 094505 (2006)  [hep-lat/0607020].

\bibitem{Fukushima:2008xe} 
  K.~Fukushima, D.~E.~Kharzeev and H.~J.~Warringa,
  Phys.\ Rev.\ D {\bf 78}, 074033 (2008)  [arXiv:0808.3382 [hep-ph]].

\bibitem{Hasenfratz:1983ba} 
  P.~Hasenfratz and F.~Karsch,
  Phys.\ Lett.\ B {\bf 125}, 308 (1983).

\bibitem{Karsten:1980wd}
  L.~H.~Karsten and J.~Smit,
  Nucl.\ Phys.\  B {\bf 183}, 103 (1981).

\end{thebibliography}
\end{document}